\documentclass[journal]{IEEEtran}

\usepackage{graphicx}
\usepackage{color}
\usepackage{algorithm}
\usepackage{psfrag}
\usepackage{subfigure}
\usepackage{amsthm}
\usepackage{url}
\usepackage{stfloats}
\usepackage{amssymb, amsfonts, amsmath}
\usepackage{fancyhdr}

\makeatletter
    
    \newcommand{\Rmnum}[1]{\expandafter\@slowromancap\romannumeral #1@}
  \makeatother

\usepackage{array}

\begin{document}

\title{Identifying spatial invasion of pandemics on metapopulation networks via anatomizing \\arrival history*}

\author{Jian-Bo Wang,~\IEEEmembership{Student Member,~IEEE}, Lin~Wang,~\IEEEmembership{Member,~IEEE},~and~Xiang~Li,~\IEEEmembership{Senior Member,~IEEE}

\thanks{This work was partially supported by the National Science Fund for Distinguished Young Scholars of China (No.~61425019),
the National Natural Science Foundation (No.~61273223), and Shanghai SMEC-EDF Shuguang project (14SG03). J.B.W. and L.W. also
acknowledge the partial support from Fudan University Excellent Doctoral Research Program (985 Program).}

\thanks{J.-B. Wang and X. Li are with the Adaptive Networks and Control Laboratory, Department of Electronic Engineering, and
with the Research Center of Smart Networks and Systems, School of Information Science and Engineering,
Fudan University, Shanghai 200433, China (e-mail: \{jianbowang11, lix\}@fudan.edu.cn).}

\thanks{L. Wang was with the Adaptive Networks and Control Laboratory, Department of Electronic Engineering, School of Information Science and Engineering, Fudan University, Shanghai 200433, China, and is with the School of Public Health, Li Ka Shing Faculty of Medicine, The University of Hong Kong, Hong Kong SAR,
China (email: sph.linwang@hku.hk).}

\thanks{* All correspondences should contact X. Li.}}
\maketitle

\thispagestyle{fancy}
\fancyhead{}
\lhead{}
\chead{}
\rhead{}
\lfoot{}
\cfoot{\thepage}
\rfoot{}
\renewcommand{\headrulewidth}{0pt}

\pagestyle{plain}
\cfoot{\thepage}

\begin{abstract}
Spatial spread of infectious diseases among populations via the mobility of humans is highly stochastic and heterogeneous. Accurate
forecast/mining of the spread process is often hard to be achieved by using statistical or mechanical models. Here we propose a new
reverse problem, which aims to identify the stochastically spatial spread process itself from observable information regarding the
arrival history of infectious cases in each subpopulation. We solved the problem by developing an efficient optimization algorithm based on dynamical
programming, which comprises three procedures: i, anatomizing the whole spread process among all subpopulations into disjoint componential
patches; ii, inferring the most probable invasion pathways underlying each patch via maximum likelihood estimation; iii, recovering
the whole process by assembling the invasion pathways in each patch iteratively, without burdens in parameter calibrations and computer
simulations. Based on the entropy theory, we introduced an identifiability measure to assess the difficulty level that an invasion pathway can be identified. Results on
both artificial and empirical metapopulation networks show the robust performance in identifying actual invasion pathways driving pandemic spread.
\end{abstract}

\begin{IEEEkeywords}
Spatial spread, infectious diseases, metapopulation, networks, process identification, identifiability.
\end{IEEEkeywords}

\section{Introduction}
\IEEEPARstart {T}{he} frequent outbreaks of emerging infectious diseases in recent decades lead to great social, economic, and public health burdens \cite{KMJRP}-\cite{Fitch:15}. This trend is partially due to the urbanization process and, in particular, the establishment of long-distance traffic networks, which facilitate the dissemination of pathogens accompanied with passengers \cite{Dirk:13,McMichael:13}. Real-world examples include the trans-national spread of SARS-CoV in 2003 \cite{MMPW06SARS}, the global outbreak of A (H1N1) pandemic flu in 2009 \cite{Science3241557,Science326729}, avian influenza in southeast Asia \cite{YuHJ13Lancet,Lam:15}, the spark of Ebola infections in western countries in 2014 \cite{Gomes:14}, and recent potential outbreak of MERS \cite{Cowling:15}.

During almost the same epoch, the theory of complex networks has been developed as a valuable tool for modeling the structure and dynamics of/on complex systems \cite{Wang:03}-\cite{Yang:12}. In the study of network epidemiology, networks are often used to describe the epidemic spreading from human to human via contacts, where nodes represent persons and edges represent interpersonal contacts \cite{Fu:13}-\cite{Chen:04}. To characterize the spatial spread between different geo-locations, simple network models are generalized with metapopulation framework, in which each node represents a population of individuals that reside at the same geo-region (e.g. a city), and the edge describes the traffic route that drives the individual mobility between populations \cite{Pastor:15}, \cite{WL14CSB}. The networked metapopulation models have been applied to study the real-world cases such as SARS \cite{MMPW06SARS}, A (H1N1) pandemic flu \cite{Balcan:09}, Ebola \cite{Gomes:14}, which can capture some key dynamic features including peak times, basic epidemic curves, and epidemic sizes. Quantitative model results can be used to evaluate the effectiveness of control strategies \cite{Ferguson06Nature}-\cite{Wu09PlosMed}, such as optimizing the vaccine allocation.

The numerical computing of large-scale metapopulation models is time-consuming, because of the requirement of high-level computer power. The model calibrations need high-resolution data for incidence cases, which may not be available or accurate during the early weeks of initial outbreaks \cite{Dirk:13}. Hence, continuous model training with data collected in real-time is essential in achieving a reliable model prediction \cite{Lofgren:14}. Generally, model results are the ensemble average over numerous simulation realizations, which aims to predict the mean and variance of epidemic curves, while in reality there is no such thing described by the average over different realizations \cite{Gau:08}. To
extract more meaningful information from epidemic data generated by surveillance systems, recent studies (particularly in engineering fields)
start paying attention to reverse problems, such as source detection and network reconstruction, which are briefly summarized here.

\subsection{Related Works}
The theory of system identification has been established in engineering fields, usually used to infer system parameters. The use of system identification in epidemiology mainly focuses on inferring epidemic parameters, such as the transmission rate and generation time~\cite{Miao:11},
which relies on constructing dynamical systems of ordinary differential equations. The methodology of system identification is not helpful
in solving high-dimensional stochastic many-body systems, such as metapopulation models.

Source detection for rumor spreading on complex networks is becoming a popular topic, attracting extensive discussions in recent years. The
target is to figure out the causality that can trigger the explosive dissemination across social networks, such as Facebook, Twitter, and
Weibo. For example, using maximum likelihood estimators, D. Shah and T. Zaman~\cite{Shah:11} proposed the concept of rumor centrality that
quantifies the role of nodes in network spreading. W. Luo et al.~\cite{Luo:13a} designed new estimators to infer infection sources and regions in large networks. Z. Wang et al.~\cite{Wangz:14}-\cite{Wangz:15} extended the scope by using multiple observations, which largely improves the detection accuracy. Another interesting topic is the network inference, which engages in revealing the topology structure of a network from the hint underlying the dynamics on a network \cite{Han:15}. Some useful algorithms (e.g. NetInf) have been proposed in refs.~\cite{Gomez:10}-\cite{Danes:14}. Note that the algorithms for source detection and network inference are not feasible in identifying
the spreading processes on metapopulation networks.

Using metapopulation networks models, some heuristic measures have been proposed to understand the spatial spread of infectious diseases, which are most related to this work. Gautreau et al. \cite{Gau:08} developed an approximation for the mean first arrival time between populations that have direct connection, which can be used to construct the shortest path tree (SPT) that characterizes the average transmission pathways among populations. Brockmann et al.~\cite{Dirk:13} proposed a measure called `effective distance', which can also be used to build the SPT.
Using a different method based on the maximum likelihood, Balcan et al.~\cite{Duy:09} generated the transmission pathways by extracting the minimum spanning tree from extensive Monte Carlo simulation results. Details about these measures will be given in Sec. IV, which compares the algorithmic performance.

\subsection{Motivation}
Current algorithms to inferring pandemic spatial spread generally make use of the topology features of metapopulation networks or extensive epidemic simulations. The resulting outcome is an ensemble average over all possible transmission pathways, which may fail in capturing those indeed transmitting the disease between populations, because of the high-level stochasticity and heterogeneity in the spreading process.

Good news comes from the development of modern sentinel and internet-based surveillance systems, which becomes increasingly popular in guiding public health control strategies. Such systems can or will provide high-resolution, location-specific data on
human and poultry cases \cite{Kwok:13}. Human mobility data are also available from mass transportation systems or GPS-based mobile Apps \cite{Fitch:15}.  Integrating these data often used in different fields, a natural reverse problem poses itself, which is the central interest of this work: Is it probable to design an efficient algorithm to identify or retrospect the stochastic pandemic spatial spread process among populations by linking epidemic data and models?

\subsection{Our Contributions}

\noindent Main contributions of this work are as follows:


\noindent i) A novel reverse problem of identifying the stochastic pandemic spatial spread process on metapopulation networks is proposed, which cannot be solved by existing techniques.

\noindent ii) An efficient algorithm based on dynamical programming is proposed to solve the problem, which comprises three procedures.
Firstly, the whole spread process among all populations will be decomposed into disjoint componential patches, which can be categorized
into four types of invasion cases. Then, since two types of invasion cases contain hidden pathways, an optimization approach based on the
maximum likelihood estimation is developed to infer the most probable invasion pathways underlying each path. Finally, the whole spread
process will be recovered by assembling the invasion pathways of each patch chronologically, without burdens in parameter calibrations
and computer simulations.

\noindent iii) An entropy-based measure called \emph{identifiability} is introduced to depict the difficulty level an invasion case can be identified. Comparisons on
both artificial and empirical networks show that our algorithm outperforms the existing methods in accuracy and robustness.




The remaining sections are organized as follows: Sec.~\ref{sec.2} provides the preliminary definitions and problem formulation; Sec.~\ref{sec.3}
describes the procedures of our identification algorithm, and introduces the identifiability measure; Sec.~\ref{sec.4} performs computer experiments
to compare the performance of algorithms; and Sec.~\ref{sec.5} gives the conclusion and discussion.

\section{Preliminary and problem formulation}\label{sec.2}
This section first elucidates the structure of networked metapopulation model, and then provides the preliminary definitions and problem
formulation.

\subsection{Networked Metapopulation Model}

In the networked metapopulation model, individuals are organized into social units such as counties and cities, defined as subpopulations,
which are interconnected by traffic networks of transportation routes. The disease prevails in each subpopulation due to interpersonal
contacts, and spreads between subpopulations via the mobility of infected persons. Fig. 1 illustrates the model structure.

\begin{figure}[!t]
\centering
\includegraphics[width=3.4in]{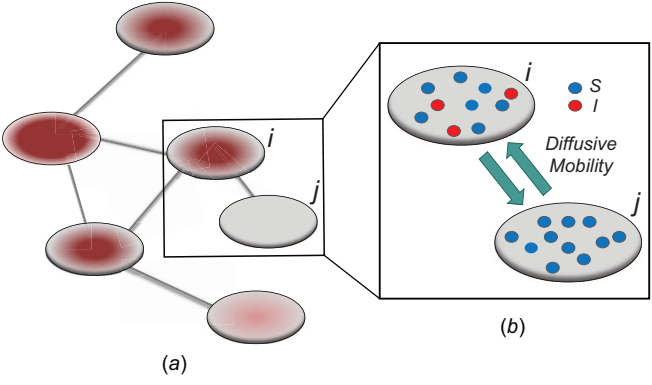}
\caption{Illustration of a networked metapopulation model, which comprises six subpopulations/patches that are coupled by the mobility of
individuals. In each subpopulation, each individual can be in one of two disease statuses (i.e. susceptible and infectious), shown in
different colors. Each individual can travel between connected subpopulation. (a) Networked metapopulation. (b) two subpopulations.}
\label{fig_sim}
\end{figure}

Within each subpopulation, individuals mix homogeneously. This assumption is partially supported by recent empirical findings on intra-urban
human mobility patterns~\cite{WL14CSB}, \cite{MB10PR}-\cite{LZD13SR}. The intra-population epidemic dynamics are characterized by compartment
models. Considering the wide applications in describing the spread of pathogens, species, rumors, emotion, behavior, crisis, etc. \cite{Shah:11}, \cite{Luo:13a}, \cite{Dong:13}, {\cite{Bro:14}}, we used the susceptible-infected (SI) model in this work. Define $N_i$ as the population size
of each subpopulation $i$, $I_i(t)$ the number of infected cases in subpopulation $i$ at time $t$, $\beta$ the transmission rate that an infected
host infects a susceptible individual shared the same location in unit time. As such, the risk of infection within subpopulation $i$ at time $t$
is characterized by $\lambda_i(t) = \beta I_i(t)/N_i$. Per unit time, the number of individuals newly infected in subpopulation $i$ can be
calculated from a binomial distribution with probability $\lambda_i(t)$ and trails equalling the number of susceptible persons $S_i(t)$.


The mobility of individuals among subpopulations is conceptually described by diffusion dynamics, $\partial_t\mathcal {X}_i=\sum_{j\in\nu(i)}p_{ji}\mathcal {X}_{j}(t)-p_{ij}\mathcal {X}_{i}(t)$, where $\mathcal {X}_i(t)$ is a placeholder for $S_i(t)$ or $I_i(t)$, $\nu(i)$ is the set of subpopulations directly connected with subpopulation $i$, and $p_{ij}$ is the per capita mobility rate from subpopulation $i$ to $j$, which equals the ratio
between the daily flux of passengers from subpopulation $i$ to $j$ and the population size of departure subpopulation $i$. The ensemble of
mobility rates $0\leq p_{ij}<1$ defines a transition matrix $\mathcal {P}$, determined by the topology structure and traffic fluxes of the
mobility network. The inter-population mobility of individuals is simulated with binomial or multinomial process (Appendix A). More details in modeling
rules can refer to our review paper~\cite{WL14CSB}.




\begin{figure*}[!t]
\centering
\includegraphics[width=4.5in]{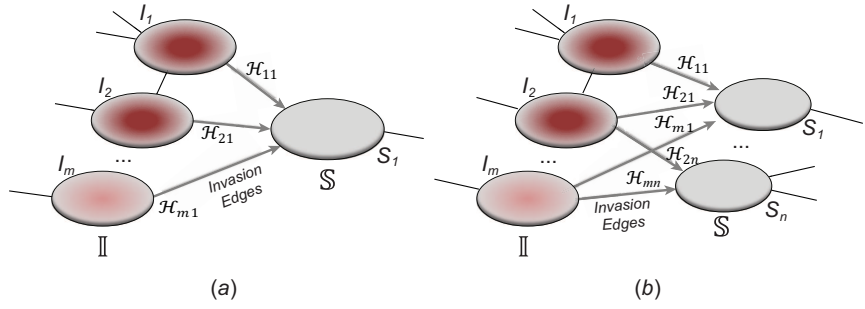}
\caption{ (a) An example of the $mI\mapsto S$ invasion case, in which $m$ infected subpopulations invade one susceptible subpopulation. The red patches denote the infected subpopulations, while the plain patch is the subpopulation that remains susceptible before
time $t$ but will be contaminated during that time step due to the arrival of infectious cases from upstream infected subpopulations. (b) An example of the $mI\mapsto nS$ invasion case, in which $m$ infected subpopulations invade $n(n\geq 2)$ susceptible subpopulations.}
\label{mInS}
\end{figure*}

\subsection{Basic Definitions}

The {\it epidemic arrival time} (EAT) is the first arrival time of infectious hosts traveling to a susceptible subpopulation.
At a given EAT, at least an unaffected (susceptible) subpopulation will be contaminated, characterizing the occurrence of {\it invasion event(s)}. Herein, $S(I)$ denotes a(an) susceptible(infected) subpopulation.

For an invasion event, organizing newly contaminated subpopulations (remaining unaffected prior to that invasion event) into set $\mathbb S$, and
infected subpopulations into set $\mathbb I$, we define the four types of {\it invasion case}~(INC) as follows:

\noindent (i) $I\mapsto S$: $\mathbb I$ and $\mathbb S$ both are composed of a single subpopulation respectively, which represents that a
previously unaffected subpopulation is infected by the new arrival of infectious host(s) from its unique neighboring infected subpopulation.

\noindent (ii) $I\mapsto nS (n>1)$: In this case, $\mathbb I$ only consists of a single subpopulation, while $\mathbb S$ contains
$n (n>1)$ subpopulations. This represents that $n$ previously unaffected subpopulations are contaminated due to the new arrival of
infectious hosts from their common infected subpopulation in $\mathbb I$.

\noindent (iii) $mI\mapsto S (m>1)$: $\mathbb S$ only consists of a single subpopulation, and $\mathbb I$ contains $m (m>1)$ subpopulations.
This means that the newly infected subpopulation in $\mathbb S$ is infected by the arrival of infected host(s) from
$m$ potential upstream subpopulations in $\mathbb I$ through the invasion edges.

\noindent (iv) $mI\mapsto nS (m,n>1)$: In this case, $\mathbb S$ and $\mathbb I$ both are composed of no less than two subpopulations,
and they constitute a connected subgraph. Each previously unaffected subpopulation in $\mathbb S$ is contaminated due to
the simultaneous arrival of infected hosts from $m$ potential source subpopulations in $\mathbb I$. Each subpopulation in $\mathbb I$ may lead to
the contamination of at least one but no more than $n$ neighboring downstream subpopulations in $\mathbb S$ through the invasion edges.
Multiple edges between any pair of subpopulations are forbidden.

Figure~\ref{mInS}(a)--(b) illustrate the two scenarios of $mI\mapsto S (m>1)$ and $mI\mapsto nS (m,n>1)$. A decomposition procedure of {\em invasion partition} (INP) is used to generate the components of invasion cases in each invasion event. The heuristic search algorithm to proceed the invasion partition is given in Algorithm I if an invasion event occurs.

\subsection {Problem Formulation}
 Suppose that the spread starts at an infected subpopulation. It forms the invasion pathways when this source invades many susceptible subpopulations and the cascading invasion goes on. We record the infected individuals of each subpopulation per unit time. From the data, we should know when a subpopulation is infected and how many infected individuals in this subpopulation, but we may not know which infected subpopulations invade this subpopulation if it has ($m$, $m\geq2$) infected neighbor subpopulations through the corresponding edge(s) (see Figure~\ref{mInS}(a)) at that time step. The question of interest is how to identify the instantaneous spatial invasion process just according to the surveillance data. Herein, we know the network topology including subpopulation size and travel flows, such as the city populations of airports and travelers by an airline of the real network of American airports network.

Define an invasion pathway which are the directed edges that infected individuals invade to susceptible subpopulations at EAT. To identify it, we proceed the following invasion pathways identification(IPI) algorithm: \\
i) Decompose the whole pathways as four types of invasion cases by the invasion partition at each EAT; Suppose the whole invasion pathways $\mathrm{T}$ are anatomized into $\Lambda$ of four invasion cases. Let $\hat{a}_i$ denote the identified invasion pathways based on the surveillance data $G$ of that invasion case $i$ and the given graph $G$. According to the (stochastic) dynamic programming, we have the following equation to optimally solve this problem

\begin{eqnarray}
\mathrm{T}_{\textrm{whole\ invasion\ pathways}} = \textrm{opt}\sum_{i=1}^\Lambda \hat{a}_i.
\end{eqnarray}

ii) For each invasion case, we first judge whether it has a unique set of invasion pathways or more than one potential invasion pathways. When an invasion case has more than one possible invasion pathway, each set of which is called potential invasion pathway. If it has more than one potential invasion pathway, we estimate the true invasion pathways $a^*_i$, denoted by $\hat{a}_i$, based on the surveillance data $G$ of that invasion case and the given graph $G$. A potential pathway belonged to that invasion case is denoted by $\forall a_i \in $ $G_{INC_i}$. To make this estimation, we shall compute the likelihood of a potential invasion pathway $a_i$. With respect to this setting, the maximum likelihood (ML) estimator of $a^*_i$ with respect to the networked metapopulation model given by that invasion case maximizes the correct identification probability. Therefore, we define the ML estimator
\begin{eqnarray}
\hat{a}_i = \mathop{\arg \max}\limits_{a_i\in G_{INC_i}}~P(a_i|G_{INC_i}),
\end{eqnarray}
where $P(a_i|G_{INC_i})$ is the likelihood of observing the potential pathway $a_i$ assuming it's the true pathway $a^*_i$.
Thus we would like to evaluate $P(a_i|G_{INC_i})$ for all $a_i\in G_{INC_i}$ and then choose the maximal one.

\begin{algorithm}
\caption{Invasion Partition}
\label{alg:framwork}
1:  for an invasion event, collect all newly infected $S$ as initially $\mathbb S$ and their previously infected neighbors as $\mathbb I$;\\
2:  start with an arbitrary element $S_i$ in set $\mathbb S$;\\
3:  find all neighbors $\mathbb I^*$ of $S_i$ in set $\mathbb I$;\\
4:  find the new neighbors $\mathbb S^*$ in the $\mathbb S$ if have;\\
5:  find the new neighbors in the $\mathbb I$ if have;\\
6:  repeat the above two steps until cannot find any new neighbors in $\mathbb S$ and $\mathbb I$, we get an invasion case consisting of $\mathbb I^*$ and $\mathbb S^*$, then update the $\mathbb S$ and $\mathbb I$;\\
7:  repeat the 2-6 steps to get new invasion cases until there are no elements in $\mathbb S$.\\
\end{algorithm}

\section{identification algorithm to Invasion pathway}\label{sec.3}
According to our above invasion partition decompose algorithm, it is easy to identify the invasion pathways for the invasion case scenario $I\mapsto nS (n\geq 1)$ (they have the only invasion pathway from their neighbor infected subpopulation). Thus our invasion pathways' identification algorithm mainly deals with the other two kinds of invasion cases $mI\mapsto S (m>1)$ and $mI\mapsto nS (m,n>1)$. To make the description clear, we restate the term $I_i$ denotes subpopulation $i$ which is infected, and its number of infected individuals of $I_i$ at time $t$ is denoted by $I_i(t)$.

\begin{figure}[!t]
\centering
\includegraphics[width=3.45in]{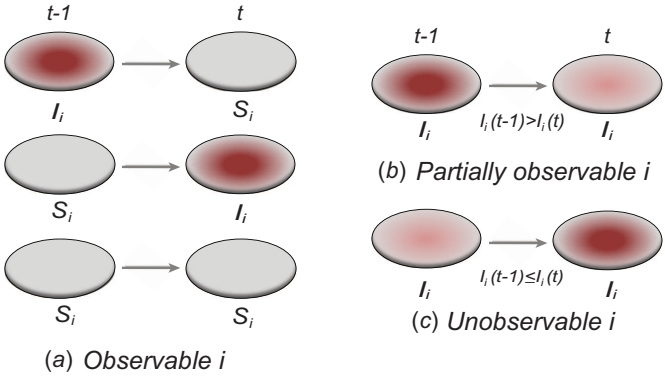}
\caption{Illustration of neighbors classification in terms of status transitions: i) unobservable subpopulations ii) partial unobservable subpopulations, and iii) observable subpopulations.}
\label{fob}
\end{figure}

As time evolves, infected hosts travel among subpopulations, inducing the spatial pandemic dispersal. For each invasion case, by
analyzing the variance of infected hosts in each subpopulation $i$, we define three levels of extent of subpopulations observability to reflect the information
held for the inference of relevant invasion pathway:

\noindent (i) {\it Observable Subpopulation:} Subpopulation $i$ is observable during an invasion case, given the occurrence of the three most evident (subpopulation's) status transitions. The first refers to the transition $S_i\rightarrow I_i$, accounting that the previously
unaffected subpopulation $i$ is contaminated during that invasion case due to the arrival of infected hosts. The second concerns the
transition $I_i\rightarrow S_i$, in which the previously infected subpopulation $i$ becomes susceptible again during that invasion case,
since the infected hosts do not trigger a local outbreak and leave $i$. In the third transition $S_i\rightarrow S_i$, despite of
having infected subpopulations in the neighborhood, subpopulation $i$ remains unaffected during that invasion case due to no arrival of
infected hosts. Figure~\ref{fob}(a) illustrates such observable transitions.

\noindent (ii) {\it Partially Observable Subpopulation:} Subpopulation $i$ is partially observable during an invasion case occurring at time $t$, if its number
of infected hosts is decreased, i.e., $I_i(t)<I_i(t-1)$ and $I_i(t)>0$, which implies that at least $\Delta I_i(t)=|I_i(t)-I_i(t-1)|$
infected hosts leave $i$ during that invasion case. It is impossible to distinguish their mobility destinations unless the invasion case $I\mapsto S$
or $I\mapsto nS$ occurs. Fig.~\ref{fob}(b) illustrates the partially observable subpopulation.

\noindent (iii) {\it Unobservable Subpopulation}: Subpopulation $i$ is unobservable during an INC occurring at time $t$, if its number of infected
hosts has not been decreased, i.e., $I_i(t)\geq I_i(t-1)$, considering the difficulty in judging whether there present infected hosts
leaving subpopulation $i$ during that invasion case. See Fig.~\ref{fob}(c) for an illustration.

We further categorize the edges emanated from each infected subpopulation
in set $\mathbb I$ into four types, i.e., invasion edges, observable edges, partially observable edges and unobservable edges:

\noindent (i) {\it Invasion Edges:} In an invasion case, invasion edges represent each route emanated from subpopulation $i$ in $\mathbb I$ to subpopulation $j$ in $\mathbb S$. They are considered as a unique category, because invasion edges contain all invasion pathway (an invasion pathway must be an invasion edge, but an invasion edge may not an invasion pathway). In Figure \ref{mInS}(a)-(b), the invasion edges are
illustrated. The following three types of edges are not belong to the routes between sets $\mathbb I$ and $\mathbb S$, but they are the edges emanated from $i$ to subpopulation $j$ that is not belong to $\mathbb S$.

\noindent (ii) {\it Observable Edges:} For infected subpopulation $i$ in $\mathbb I$, any edge emanated from $i$ is observable, if it connects $i$
with observable subpopulation $j$ that only experiences the transition $S_j\rightarrow S_j$ or $I_j\rightarrow S_j$ from $t_{EAT-1}$ to $t_{EAT}$. Here, it
is intuitive that in subpopulation $j$ there is no arrival of infected hosts from subpopulation $i$.

\noindent (iii) {\it Partially Observable Edges:} For infected subpopulation $i$ in $\mathbb I$, any edge is partially observable, if it connects $i$ with a partially observable subpopulation.

\noindent (iv) {\it Unobservable Edges:} For infected subpopulation $i$ in $\mathbb I$, any edge is unobservable, if it connects $i$ with an unobservable subpopulation.

The classification of subpopulations and edges are used to compute the corresponding subpopulation's transferring estimator in the following section III of both invasion cases of $mI\mapsto S (m>1)$ and $mI\mapsto S (m,n>1)$.

\subsection{The Case of $mI\mapsto S (m>1)$}

As shown in Figure \ref{mInS}(a), a
typical INC $mI\mapsto S (m>1)$ is composed of two sets of subpopulations, i.e., the previously infected subpopulations $\mathbb I=\{I_1,I_2,\ldots,I_m\}$ and the previously unaffected subpopulation $\mathbb S=\{S_1\}$. Suppose that subpopulation $S_1$ is contaminated at time $t$ due to the appearance of $\mathcal {H}$ infected hosts ($\mathcal {H}$ is a positive integer number) that
come from the potential sources in $\mathbb I$. If the actual number of infected hosts from subpopulation $I_i$ is $\mathcal {H}_{i1}$, $i\in\mathbb I$, we have
\begin{equation}
\sum^{m}_{i=1}\mathcal {H}_{i1}=\mathcal {H},\label{eqh}
\end{equation}
with the conditions $0\leq \mathcal {H}_{i1}\leq\mathcal {H}$ and $\mathcal {H}_{i1}\leq I_i(t-1)$.

\noindent (i) {\it  Accurate Identification of Invasion Pathway}

Given a few satisfied prerequisites,
Eq. (\ref{eqh}) can has a unique solution, which implies that the invasion pathways of that invasion case can be identified accurately.
Theorem~1 elucidates this scenario.

\noindent {\em Theorem 1 (Accurate Identification of Invasion Pathway):} With the following conditions: 1) Among $m$ possible sources illustrated in set $\mathbb I$, there are only $m' (m' \leq m)$ partially observable subpopulations $\mathbb I'$, whose neighboring subpopulations (excluding the invasion destination $S_1$) only experience the transition
$S$ to $S$ or $I$ to $S$ at that EAT, 2) $\sum_{i\in \mathbb I'} \big[I_i(t-1) - I_i(t)\big] = \mathcal {H}$, the invasion pathway of an invasion case
$mI\mapsto S (m>1)$ can be identified accurately.
\begin{IEEEproof}
According to the definition of observability, in an INC, the number of local infected hosts in an involved partially observable
source $i$ will be decreased by \big[$I_i(t-1)-I_i(t)$\big] due to their departure. If the subpopulations in the neighborhood
of $i$ only experience the transition of $S_i$ to $S_i$ or $I_i$ to $S_i$ from $t_{EAT-1}$ to $t_{EAT}$, they are impossible to receive the infected
hosts from subpopulation $i$. Therefore, the newly contaminated subpopulation $S_1$ is the only destination for those infected
travelers departing from the partially observable sources. Since $m'\leq m$, the second condition guarantees that Eq. (\ref{eqh})
only has a unique solution, which corresponds to the accurate identification of invasion pathways of this invasion case.
\end{IEEEproof}
\noindent (ii) {\it Potential Invasion Pathway}\\
If the conditions of Theorem~1 are unsatisfied, Eq. (\ref{eqh}) has multiple solutions, each solution corresponds to a set of potential
invasion pathways that can result in the related invasion case. Due to the heterogeneity in the traffic flow on each edge and the number
of infected hosts within each contaminated source, each set of potential pathways is associated with a unique likelihood, which
also identifies the occurrence probability of the corresponding solution of Eq. (\ref{eqh}). Therefore, the identification of
invasion pathway that induce an invasion case can be transformed to searching the most probable solution of Eq.~(\ref{eqh}).

We define the solution space $\Phi$ of Eq.~(\ref{eqh}) of the invasion case $mI\mapsto S (m>1)$, which subjects to two conditions: (i) $\sum^{m}_{i=1}\mathcal {H}_{i1}=\mathcal {H}$; (ii) $\forall\mathcal {H}_{i1}$, $\mathcal {H}_{i1}\leq I_i(t-1)$. The second condition is obvious, since the number of
infected travelers departing from the source $I_i$ cannot exceed $I_i(t-1)$. Let us assume that $\Phi$ contains $M$ solutions, and a
typical solution is formulated as $\sigma_{j}=\{\mathcal {H}^{(j)}_{i1}, i=[1,...,m]\}$. Obviously, each solution $\sigma_j$ corresponds to a potential invasion pathway $a_j$.

Through the invasion case $mI\mapsto S (m>1)$, the observed event $\mathcal {E}$ shows that the destination $S_1$ is contaminated
due to the arrival of totally $\mathcal {H}$ infected hosts from the potential sources $I_i, i=[1,...,m]$. With this posterior
information, we first measure the likelihood of each possible solution $\sigma_{j}$, which corresponds to the reasoning event
that for each source $I_i$, $i\in[1,m]$, $\mathcal {H}_{i1}$ infected hosts are transferred to $S_1$. It is evident that
$\forall j, P(\mathcal {E}_{mIS}|\sigma_{j})=1$, since $\sigma_{j}$ will lead to the occurrence of event $\mathcal {E}_{mIS}$, which corresponds to $G_{mIS}$.

According to Bayes' theorem, the likelihood of the solution $\sigma_{j}$ is characterized by
\begin{equation}
\begin{aligned}
P(\sigma_j|\mathcal {E}_{mIS})&=P(\mathcal {E}_{mIS}|\sigma_j)P(\sigma_j)\big/P(\mathcal {E}_{mIS})\\
&=P(\mathcal {E}_{mIS}|\sigma_j)P(\sigma_j)\Big/\sum_{j=1}^{M}\big[P(\mathcal {E}_{mIS}|\sigma_j)P(\sigma_j)\big]\\
&=P(\sigma_j)\Big/\sum_{j=1}^{M}\big[P(\sigma_j)\big]\\
&=\prod_{k=1}^m \Omega(\mathcal {H}^{(j)}_{k1})\Big/\sum_{i=1}^M \prod_{k=1}^m\Omega(\mathcal {H}^{(i)}_{k1}),
\end{aligned}
\label{eqMTO}
\end{equation}
where $M$ represents the number of potential solution $\sigma_j$, and the last item $\Omega(\mathcal {H}^{(i)}_{k1})$ represents the mobility likelihood transferring estimator of infected subpopulation $I_k$ in $\mathbb I$.

One linchpin of our algorithm in handling the scenario $mI\mapsto S (m>1)$ is to estimate the probability
of transferring $\mathcal {H}_{i1}$ infected hosts from each infected subpopulation $I_i, i\in \mathbb I,$ to the destination
subpopulation $S_1$. Based on the independence between the intra-subpopulation epidemic reactions and the inter-subpopulation
personal diffusion, we introduce a \emph{transferring estimator} to analyze the individual mobility of each source $I_i$, which is
in particular useful if there are partially observable and unobservable edges emanated from the focal infected subpopulation.

The specific formalisms of the transferring estimator are defined according to the three types of infected subpopulation $I_i$ consisted of set $\mathbb I$ which are unobservable subpopulation, partially unobservable subpopulation and observable subpopulation with transition of $I$ to $S$.

\noindent \emph{Unobservable Subpopulation $I_i$:} Due to the occurrence of $mI\mapsto S$, among all $k_i$ edges emanated from
subpopulation $I_i$, there is only one invasion edge in that invasion case, labeled as $k_i$, along which the traveling rate is $p_{k_i}$ and
$\mathcal {H}_{i1}$ infected hosts are transferred to the destination $S_1$. Assume that there are $\ell_i$ ($1\leq\ell_i< k_i$)
unobservable and partially unobservable edges, labeled as $1,2,...,\ell_i$, respectively. Along each unobservable or partially
unobservable edge, the traveling rate is $p_{\ell}$, $\ell\in[1,\ell_i]$, and $x_{\ell}$ infected hosts leave $I_i$. Accordingly,
in total $\eta_i = \sum_{\ell}x_{\ell}$ infected hosts leave $I_i$ through the unobservable and partially unobservable edges.
There remain $k_i-\ell_i-1$ observable edges, labeled as $\ell_i+1,...,k_i-1$, respectively. Along each observable edge, the
traveling rate is $p_\aleph$, $\aleph\in[\ell_i+1,k_i-1]$, and $x_\aleph$ infected hosts leave $I_i$. With probability
$\overline{p_{i}}=1-p_{k_i}-\sum_{\ell}p_{\ell}-\sum_\aleph p_\aleph$, an infected host keeps staying at source $I_i$.

Since the infected hosts transferred by unobservable and partially unobservable edges are untraceable, it is unable
to reveal the actual invasion pathways resulting in that invasion case accurately. Fortunately, the message of traveling rates on
each edge is available by collecting and analyzing the human mobility transportation networks. Therefore, the mobility multinomial
distribution (Appendix A Eq. (\ref{eq.2})) can be used to obtain the conditional probability that $\mathcal {H}_{i1}$ infected hosts are
transferred from infected source $I_i$ to destination $S_1$, which is measured by the following transferring estimator:
\begin{eqnarray}
\Omega_u(\mathcal {H}_{i1})&=P\big(\mathcal {H}_{i1},p_{k_i};I_i(t-1);x_\ell,p_\ell,\ell=[1,...,\ell_i];\nonumber\\
&x_\aleph,p_\aleph,
\aleph=[\ell_i+1,...,k_i-1];\overline{x_i},\overline{p_i}\big),
\end{eqnarray}
where $\overline{x_i}$ accounts for the number of infected hosts that do not leave source $I_i$ after the invasion case. Here, the
observed number of infected persons in source $I_i$ before the invasion case, i.e., $I_i(t-1)$, is used for the estimation, since
the probability that a newly infected host also experiencing the mobility process is very low. Considering the conservation
of infected hosts, and the implication of observable edges (i.e., $x_\aleph=0, \forall \aleph$), we have
$I_i(t-1)=\mathcal {H}_{i1}+\sum_{\ell}x_{\ell}+\overline{x_i}=\mathcal {H}_{i1}+\eta_i+\overline{x_i}$.
Taking into account all scenarios that fulfill the condition $\eta'_i=\eta_i+\overline{x_i}=I_i(t-1)-\mathcal {H}_{i1}$, the
transferring estimator is simplified by the marginal distribution of Eq.(4), i.e.,
\begin{equation}
\begin{aligned}
&\mathop{\sum}\limits_{\eta'_i=I_i(t-1)-\mathcal {H}_{i1}}P\Big(\overline{x_i},x_\ell,\ell=[1,...,\ell_i]\Big)=\\
&\mathop{\sum}\limits_{\eta'_i=I_i(t-1)-\mathcal {H}_{i1}}\frac{I_i(t-1)!}{\mathcal {H}_{i1}!\prod_\ell x_\ell!\overline{x_i}!}
p^{\mathcal {H}_{i1}}_{k_i}\prod_\ell p^{x_\ell}_{\ell}\overline{p_i}^{\overline{x_i}}.
\end{aligned}
\end{equation}
With independence, the transferring estimator becomes
\begin{eqnarray}
\Omega_u=\frac{I_i(t-1)!}{\mathcal {H}_{i1}!\eta'_i!}p^{\mathcal {H}_{i1}}_{k_i}\Big[\sum_\ell p_{\ell}+\overline{p_i}\Big]^{\eta'_i}.
\end{eqnarray}

\noindent {\em Observable Subpopulation $I_i$ ($I_i$ to $S_i$):} If the infected hosts of source $I_i$ all leave to travel from $t_{EAT-1}$ to $t_{EAT}$, the subpopulation $I_i$
is observable at that invasion case. In this case, we have additional posterior messages, i.e., $I(t)=0$, $\Delta I_i(t) = I(t-1)$. Here,
the number of infected hosts transferred to $S_1$ cannot exceed the total number of infected travelers departing from
source $I_i$, i.e., $\mathcal {H}_{i1}\leq \Delta I_i(t)$. In this regard, the probability that $\mathcal {H}_{i1}$ infected
hosts arrive in destination $S_1$ is measured by the following transferring estimator
\begin{equation}
\dbinom{\Delta I_i(t)}{\mathcal {H}_{i1}}
\Big[\frac{p_{k_i}}{\sum_\ell p_{\ell}+p_{k_i}}\Big]^{\mathcal {H}_{i1}}
\Big[1-\frac{p_{k_i}}{\sum_\ell p_{\ell}+p_{k_i}}\Big]^{\big[\Delta I_i(t)-\mathcal {H}_{i1}\big]}.
\end{equation}

\noindent \emph{Partially Observable Subpopulation $I_i$:} If source $I_i$ is partially observable, we can develop the inference
algorithm with an additional posterior message, which reveals that at least $\Delta I_i(t)=I_i(t)-I_i(t-1)\geq 1$ infected hosts
leave the focal source $I_i$ after the occurrence of that invasion case. In order to measure the conditional probability that $\mathcal
{H}_{i1}$ infected hosts are transferred from source $I_i$ to destination $S_1$, we inspect all possible scenarios
in detail, as follows:

\noindent Type 1: $\Delta I_i(t)\leq\mathcal {H}_{i1}$, i.e., the observed reduction in the number of infected hosts $\Delta I_i(t)$
is less than those transferred from $I_i$ to $S_1$. Here, we consider all cases that are in accordance with this condition.

If all $\Delta I_i(t)$ confirmed infected travelers are transferred from $I_i$ to $S_1$, the transferring estimator can be
used to quantify the conditional probability that the remaining $\mathcal {H}_{i1}-\Delta I_i(t)$ infected hosts concerned also
visit $S_1$, i.e.,
\begin{eqnarray}
&\Omega_p\big(\mathcal {H}_{i1}-\Delta I_i(t)|I_i(t-1)-\Delta I_i(t)\big)\nonumber\\
&=\frac{\big[I_i(t-1)-\Delta I_i(t)\big]!}{\big[\mathcal {H}_{i1}-\Delta I_i(t)\big]!\eta'_i!}
p^{[\mathcal {H}_{i1}-\Delta I_i(t)]}_{k_i}\Big[\sum_\ell p_{\ell}+\overline{p_i}\Big]^{\eta'_i}\cdot\nonumber\\
&\Big[\frac{p_{k_i}}{\sum_\ell p_{\ell}+p_{k_i}}\Big]^{\Delta I_i(t)},
\end{eqnarray}
where $p_{k_i}\big/\big[\sum_\ell p_{\ell}+p_{k_i}\big]$ represents the relative traveling rate that any person from source
$I_i$ is transferred to $S_1$, thus the last item on the right-hand side (rhs) accounts for the probability that $\Delta I_i(t)$
confirmed infected travelers all visit $S_1$.

If only a fraction of $\Delta I_i(t)$ confirmed infected travelers are transferred from $I_i$ to $S_1$, the situation
is more complicated. Assume that $\Delta I_i(t)-\phi$ ($1\leq\phi<\Delta I_i(t)$) confirmed travelers successfully come to $S_1$,
the corresponding transferring estimator becomes
\begin{eqnarray}
&\Omega_p(\mathcal {H}_{i1}-\Delta I_i(t)+\phi|I_i(t-1)-\Delta I_i(t))\nonumber\\
&=\dbinom{\Delta I_i(t)}{\phi}
\Big[\frac{p_{k_i}}{\sum_\ell p_{\ell}+p_{k_i}}\Big]^{\big[\Delta I_i(t)-\phi\big]}\Big[1-\frac{p_{k_i}}{\sum_\ell p_{\ell}+p_{k_i}}\Big]^{\phi}\nonumber\\
&\frac{\big[I_i(t-1)-\Delta I_i(t)\big]!}{\big[\mathcal {H}_{i1}-\Delta I_i(t)+\phi\big]!\eta'_i!}p^{[\mathcal {H}_{i1}-\Delta I_i(t)+\phi]}_{k_i}\Big[\sum_\ell p_{\ell}+\overline{p_i}\Big]^{\eta'_i},
\end{eqnarray}
where the first item on the r.h.s. accounts for the probability that $\Delta I_i(t)-\phi$ confirmed visitors visit $S_1$.

If all $\Delta I_i(t)$ confirmed infected travelers from $I_i$ are not transferred to $S_1$, the conditional probability
that among the remaining $I_i(t-1)-\Delta I_i(t)$ infected hosts, $\mathcal {H}_{i1}$ infected travelers are transferred to $S_1$,
which is measured by the following transferring estimator
\begin{eqnarray}
&\Omega_p(\mathcal {H}_{i1}|I_i(t-1)-\Delta I_i(t))=\frac{[I_i(t-1)-\Delta I_i(t)]!}{\mathcal {H}_{i1}!\eta'_i!}\cdot\nonumber\\
&p^{\mathcal {H}_{i1}}_{k_i}
[\sum_\ell p_{\ell}+\overline{p_i}]^{\eta'_i}
[1-\frac{p_{k_i}}{\sum_\ell p_{\ell}+p_{k_i}}]^{\Delta I_i(t)},
\end{eqnarray}
where the last item on the r.h.s. accounts for the probability that $\Delta I_i(t)$ confirmed infected travelers all do not visit $S_1$.

Taking into account all the above cases, the probability that $\mathcal {H}_{i1}$ infected hosts arrive at destination
$S_1$ is measured by the following transferring estimator
\begin{eqnarray}
\Omega_p(\mathcal {H}_{i1})=&\sum^{\Delta I_i(t)}_{\phi=0}\dbinom{\Delta I_i(t)}{\phi}\Big[\frac{p_{k_i}}{\sum_\ell p_{\ell}+p_{k_i}}\Big]^{\big[\Delta I_i(t)-\phi\big]}\cdot\nonumber\\
&\Big[1-\frac{p_{k_i}}{\sum_\ell p_{\ell}+p_{k_i}}\Big]^{\phi}\frac{\big[I_i(t-1)-\Delta I_i(t)\big]!}{\big[\mathcal {H}_{i1}-\Delta I_i(t)+\phi\big]!\eta'_i!}\cdot\nonumber\\
&p^{[\mathcal {H}_{i1}-\Delta I_i(t)+\phi]}_{k_i}\Big[\sum_\ell p_{\ell}+\overline{p_i}\Big]^{\eta'_i}.
\end{eqnarray}

\noindent Type 2: $\Delta I_i(t)>\mathcal {H}_{i1}$, i.e., the observed reduction in the number of infected hosts $\Delta I_i(t)$
exceeds the number of infected hosts transferred to $S_1$. Similar to the above analysis, we develop the transferring estimator by
considering all possible cases that are in accordance with this condition.

If $\mathcal {H}_{i1}$ infected hosts transferred to $S_1$ are all from the observable travelers $\Delta I_i(t)$,
the transferring estimator becomes
\begin{equation}
\begin{aligned}
\dbinom{\Delta I_i(t)}{\mathcal {H}_{i1}}\Big[\frac{p_{k_i}}{\sum_\ell p_{\ell}+p_{k_i}}\Big]^{\mathcal {H}_{i1}}
&\Big[1-\frac{p_{k_i}}{\sum_\ell p_{\ell}+p_{k_i}}\Big]^{\big[\Delta I_i(t)-\mathcal {H}_{i1}\big]}\cdot\\
&\Big[\sum_\ell p_{\ell}+\overline{p_i}\Big]^{\big[I_i(t)-\Delta I_i(t)\big]},
\end{aligned}
\end{equation}
where the last item accounts for the constraint that the remaining $I_i(t)-\Delta I_i(t)$ infected hosts will not
be transferred to $S_1$.

Similar to Type 1, the other two cases are: only a fraction of $\mathcal {H}_{i1}$ infected hosts transferred to $S_1$ are from the observable travelers $\Delta I_i(t)$, and $\mathcal {H}_{i1}$ infected hosts transferred to $S_1$ are all not from the observable travelers $\Delta I_i(t)$, we can also derive the transferring estimators.

Taking into account all these cases, the probability that $\mathcal {H}_{i1}$ infected hosts move to the destination
subpopulation $S_1$ is measured by the following transferring estimator
\begin{eqnarray}
\sum_{\triangle \mathcal {H}_{i1}=0}^{\mathcal {H}_{i1}}\dbinom{\Delta I_i(t)}{\Delta \mathcal {H}_{i1}}
\Big[\frac{p_{k_i}}{\sum_\ell p_{\ell}+p_{k_i}}\Big]^{\Delta \mathcal {H}_{i1}}\cdot\nonumber\\
\Big[1-\frac{p_{k_i}}{\sum_\ell p_{\ell}+p_{k_i}}\Big]^{\big[\Delta I_i(t)-\Delta\mathcal {H}_{i1}\big]}\cdot\nonumber\\
\frac{\big[I_i(t-1) - \Delta I_i(t)]!}{(\mathcal {H}_{i1} - \Delta \mathcal {H}_{i1})!\eta'!}p_{k_i}^{\mathcal {H}_{i1}
-\Delta\mathcal {H}_{i1}}\Big[\sum_\ell p_{\ell}+\overline{p_i}]^{\eta'}.
\end{eqnarray}

Generally, set $\mathbb I$ consists of the three classes of subpopulations $I_i(1\leq i\leq m)$ discussed above: unobservable subpopulation, partially unobservable subpopulation, observable subpopulation of $I \rightarrow S$. According to Eq.(4), generally each potential pathway $a_i$ corresponds to a potential solution $\sigma_i$, the most-likely invasion pathway for a $mI \mapsto S$ can be identified as\\
\begin{equation}
\begin{aligned}
\hat{a}^{mIS} &= \arg \mathop{\max}\limits_{\sigma_i}~P(\sigma_i|\mathcal {E}_{mIS})\\
                &= \arg \mathop{\max}\limits_{a_i}~P(a_i|G_{mIS}).
\end{aligned}
\end{equation}

\subsection{The Case of $mI\mapsto nS (m>1,n>1)$}
Finally, we consider the case of $mI\mapsto nS (m>1,n>1)$, which is more complicated than $mI\mapsto S$, because some infectious populations in set $\mathbb I$ may have more than one invasion edge to the corresponding susceptible subpopulations in set $\mathbb S$, and the number of elements in set $\mathbb S$ are more than one, which obey a joint probability distribution of transferring likelihood. As shown in Fig. \ref{mInS}, an invasion case $mI\mapsto nS$ includes set $\mathbb I=\{I_i|i=1,2,\ldots,m\}$ and $\mathbb S=\{S_i|i=1,2,\ldots,n\}$. The first arrival infected individuals invaded each susceptible subpopulation in set $\mathbb S$ are $\{\mathcal {H}_i|i=1,2,\ldots,n\}$, respectively. Here, denote $U_i(i=1,2,\ldots,m)$ the subset of susceptible neighbor subpopulations in set $\mathbb S$ of infected subpopulation $I_i$ , and $Y_j(j=1,2,\ldots,n)$ the subset of infected neighbor subpopulations in set $\mathbb I$ of susceptible subpopulation $S_j$.

We define $\sigma = \{\{\mathcal {H}_{i1}|i\in Y_1\},\ldots,\{\mathcal {H}_{in}|i\in Y_n\}\}$ a potential solution for the $mI\mapsto nS$, if subjects to two conditions: (i)
\begin{eqnarray}
\sum_{i\in Y_k}\mathcal {H}_{ik} = \mathcal {H}_k,
\label{eqhmtm}
\end{eqnarray}
$\mathcal {H}_{ik} \geq 0$;
(ii) For any $\mathcal {H}_{ik}$ which denotes the number of infected hosts travel to subpopulation $S_k$ from $I_i$ at $t_{EAT}$, we have $\sum_{k\in U_i}\mathcal {H}_{ik} \leq I_i(t-1)$, where $1\leq i\leq m, 1\leq k\leq n$.\\
If a $mI\mapsto nS$ has $M$ potential solutions, let $\sigma_j = \{\{\mathcal {H}_{i1}^{(j)}|i\in Y_1\},\ldots,\{\mathcal {H}_{in}^{(j)}|i\in Y_n\}\}$, $1\leq j\leq M$.

Similarly, we first discuss the directly identifiable pathway for a given $mI\mapsto nS$, then estimate the most-likely numbers of each $\mathcal {H}_{ik}$ as accurate as possible by designing our identification algorithm, since one solution of Eq.~(\ref{eqhmtm}) corresponds to one invasion pathway of an invasion case $mI\mapsto nS$.

\noindent {(i) \it Accurate Identification of Invasion Pathway}

Given a few satisfied prerequisites, for all $k\in U_i, i\in Y_k$ of the equations constituted by Eq.(\ref{eqhmtm}) can has a unique solution, which implies that the invasion pathway of that invasion case can be identified accurately.
Theorem~2 elucidates this scenario.

\noindent {\em Theorem 2 (Accurate Identification of Invasion Pathway):} With the following conditions:
1) the number of invasion edges $E_{in} \leq n+m$,
2) the neighbor subpopulations of each subpopulation in set $\mathbb I$ are with the transition $S$ to $S$ or $I$ to $S$ except their neighbor subpopulations in set $\mathbb S$ during $t_{EAT-1}$ to $t_{EAT}$,
3) $\sum_{i=1}^m \triangle I_i(t) = \sum_{k=1}^n \mathcal {H}_k$, the invasion pathway of an invasion case
$mI\mapsto nS (m,n>1)$ can be identified accurately.

\begin{IEEEproof}
Since the number of infected individuals in the partially observable subpopulation $i$ reduces at time $t$, i.e., $I_i(t)<I_i(t-1)$,
$I_i(t)>0$, it is inevitable that a few infected carries diffuse away from subpopulation $i$. Occurring the state transitions of
$S\rightarrow I, I\rightarrow S$ at time $t$, subpopulations in the neighborhood of $i$ (excluding the new contaminated subpopulation
$j$) cannot receive infected travelers. Therefore, the only possible destination for those infected travelers is subpopulation $S_j$.

The conditions $E_{in} \leq n+m$ and $\sum_{i=1}^m \triangle I_i(t) = \sum_{k=1}^n \mathcal {H}_k$ make the equations $\sum_{i\in Y_k}\mathcal {H}_{ik} = \mathcal {H}_k$ and $\sum_{k\in U_i}\mathcal {H}_{ik} = \triangle I_i(t)$ only has the unique solution $\sigma = \{\{\mathcal {H}_{i1}|i\in Y_1\},\ldots,\{\mathcal {H}_{in}|i\in Y_n\}\}$. The reason is that rank($A_{coef}$)=$E_{in}$, where $A_{coef}$ is the coefficient matrix of equations $\sum_{i\in Y_k}\mathcal {H}_{ik} = \mathcal {H}_k$ and $\sum_{k\in U_i}\mathcal {H}_{ik} = \triangle I_i(t)$. Thus the invasion pathway of this
$mI\mapsto nS (m,n>1)$ can be identified accurately.\\
\end{IEEEproof}

\noindent (ii) {\it Potential Invasion Pathway}

If the conditions of Theorem~2 are unsatisfied, the equations constituted by Eq. (\ref{eqhmtm}) has multiple solutions, each solution corresponds to a set of potential invasion pathways that can result in the related $mI\mapsto nS (m,n>1)$. We derive the transferring likelihood of each potential solution similar to case of $mI\mapsto S$. Therefore, the likelihood of solution $\sigma_{j}$ is characterized by
\begin{eqnarray}
P(\sigma_j|\mathcal {E}_{mInS})
=\prod_{k=1}^m \Omega(\mathcal {H}^{(j)}_{kk_{\hbar}})\Big/\sum_{i=1}^M \prod_{k=1}^m\Omega(\mathcal {H}^{(i)}_{kk_{\hbar}}),
\end{eqnarray}
where $M$ represents the number of solution $\sigma_j$, and the last item $\Omega(\mathcal {H}^{(i)}_{kk_{\hbar}})$ represents the transfer estimator of infected subpopulation $I_k$ in $\mathbb I$, $k_\hbar \in Y_k$.
Note that $\sigma_j$ and $\mathcal {E}_{mInS}$ correspond to a potential invasion pathway $a_j$ of $mI\mapsto nS$ and $G_{mInS}$, respectively.

Now we discuss the transferring estimator of subpopulation $I_i$ according to its extent of subpopulation observability.\\
(a) Subpopulation $I_i$ has only one neighbor (invasion edge) in set $\mathbb S$.

In this case, the transferring estimator is the same as the depicted one in $mI\mapsto S$.\\
(b) Subpopulation $I_i$ has $\rho~(\rho\geq 2)$ neighbors (invasion edges) in set $\mathbb S$.

Suppose there are totally $k_i$ edges emanate from $I_i$ which consist of the following three kinds as: (1) There are $\rho_i$ invasion edges ($2\leq \rho_i \leq n$), labeled $1,2,\ldots,\rho_i$, along which the traveling rates are $p_\hbar,\hbar\in[1,\rho_i]$, and $\mathcal {H}_{ii_{\hbar}}$ invade the subpopulations in the subset $\{Y_i=i_{\hbar}\}$, respectively; (2) There are $\ell_i$ unobservable and partially observable edges, labeled $1+\rho_i,\ldots,\ell_i+\rho_i$, respectively.
Along each unobservable or partially unobservable edge, the traveling rate is $p_{\ell}$, $\ell\in[1,\ell_i]$, and $x_{\ell}$ infected hosts leave $I_i$. Accordingly, in total $\eta_i = \sum_{\ell}x_{\ell}$ infected hosts leave $I_i$ through the unobservable and partially unobservable edges. (3) There remain $k_i-\ell_i-\rho_i$ observable edges, labeled as $\ell_i+\rho_i+1,...,k_i$, respectively. Along each observable edge, the traveling rate is $p_\aleph$, $\aleph\in[\ell_i+\rho_i+1,k_i]$, and $x_\aleph$ infected hosts leave $I_i$. With probability $\overline{p_{i}}=1-\sum_{\hbar}p_{\hbar}-\sum_{\ell}p_{\ell}-\sum_\aleph p_\aleph$, an infected host keeps staying at the source $I_i$. There are $\overline{x_i}$ infected hosts staying in subpopulation $I_i$ with the probability $\overline{p_i}$. Because $I_i$ connects the unobservable and partially observable infected subpopulations, we only know the sum $\sum_\ell x_\ell + \overline{x_i} = \eta'$.

Now we employ the following estimators to evaluate the transferring likelihood of the three categories of $I_i$.

\noindent {\em Unobservable Subpopulation $I_i$:} Because $\triangle I_i(t) = I_i(t-1) - I_i(t) \leq 0$, we don't know whether and how many infected individuals travel to which destinations. Similar to the invasion case $mI\mapsto S$, the transferring likelihood estimator of $I_i$ is
\begin{equation}
\begin{aligned}
\Omega_u(\mathcal {H}_{ii_{\hbar}})=P(\mathcal {H}_{ii_{\hbar}}, p_{\hbar}, \hbar=[1,\ldots,\rho];x_{\ell}, p_{\ell},\ell=[1+\rho,\\
\ldots,l+\rho];x_{\aleph}, p_{\aleph}, \aleph=[l+\rho+1,\ldots,k];\overline{x_i},\overline{p_i}).
\end{aligned}
\end{equation}
\noindent By means of the observable edges, the transferring estimator can be simplified as
\begin{eqnarray}
\sum_{\eta'_i=I_i(t-1)-\sum\mathcal {H}_{ii_{\hbar}}}P(\mathcal {H}_{ii_{\hbar}}, p_{\hbar}, \hbar=[1,\ldots,\rho]; x_{\ell}, p_{\ell},\nonumber\\
\ell=[1+\rho,\ldots,l+\rho];\overline{x_i},\overline{p_i}\big)\nonumber\\
=\sum_{\eta'_i=I_i(t-1)-\sum\mathcal {H}_{ii_{\hbar}}}\frac{I_i(t-1)!}{\prod_{\hbar}\mathcal {H}_{ii_{\hbar}}!
 \eta! \prod_{\ell} x_{\ell}!\overline{x_i}!}\nonumber\\
 \quad \prod_{\hbar} p_{\hbar}^{\mathcal {H}_{ii_j}}
 (\sum_{\ell}p_{\ell})^\eta
 \overline{p_i}^{\overline{x_i}}.
\end{eqnarray}
Then the transferring estimator becomes by the marginal distribution as:
\begin{equation}
\Omega_u=\frac{I_i(t-1)!}{\prod_{\hbar}\mathcal {H}_{ii_{\hbar}}!\eta'_i!}\prod_{\hbar}p^{{H}_{ii_{\hbar}}}_{\hbar}\Big[\sum_\ell p_{\ell}+\overline{p_i}\Big]^{\eta'_i}.
\end{equation}

\noindent{\em Observable Subpopulation $I_i$ ($I_i$ to $S_i$):} For this situation, $H_i = \{\mathcal {H}_{ii_\hbar}|\hbar= 1,\ldots,\rho\}$ all come from $\triangle I_i(t)$.
The transferring likelihood estimator of a $I\rightarrow S$ observable subpopulation $I_i$ is
\begin{eqnarray}
&\Omega_{ob}=\frac{\triangle I_i(t)!}{\prod_\hbar \mathcal {H}_{ii_\hbar}!
(\triangle I_i(t)- \sum_{\hbar}\mathcal {H}_{ii_\hbar})!}\prod_\hbar (\frac{p_\hbar}{\sum_{k=1}^{l+\rho} p_k})^{\mathcal {H}''_{ii_\hbar}}\cdot\nonumber\\
&(\frac{\sum_\ell p_\ell}{\sum_{j=1}^{l+\rho} p_{j}})
^{\triangle I_i(t)-\sum_{\hbar}\mathcal {H}_{ii_\hbar}},
\end{eqnarray}
where $\triangle I_i(t) = I_i(t-1)-I_i(t)=I_i(t-1)$.

\noindent{\em Partially Unobservable Subpopulation $I_i$:} Due to $\triangle I_i(t) = I_i(t-1) - I_i(t)>0$, at least $\triangle I_i(t)$ infected hosts leave source $I_i$ from $t_{EAT-1}$ to $t_{EAT}$.

We first decompose $H_i = \{\mathcal {H}_{ii_\hbar}|\hbar=1,\ldots,\rho\}$ as two subsets: $H'_i = \{\mathcal {H}'_{ii_\hbar}|\hbar=1,\ldots,\rho\}$ and $H''_i = \{\mathcal {H}''_{ii_\hbar}|\hbar=1,\ldots,\rho\}$, $\mathcal {H}'_{ii_\hbar}+\mathcal {H}''_{ii_\hbar}=\mathcal {H}_{ii_\hbar}$, where $\mathcal {H}'_{ii_\hbar}\geq 0, \mathcal {H}''_{ii_\hbar}\geq 0$. Denote $H'_i = \{\mathcal {H}'_{ii_\hbar}|\hbar=1,\ldots,\rho\}$ the infected hosts coming from $I_i(t-1)-\triangle I_i(t)$, and $H''_i = \{\mathcal {H}''_{ii_\hbar}|\hbar=1,\ldots,\rho\}$ the infected hosts coming from $\triangle I_i(t)$.
Then we analyze the transferring estimator on the following two types.\\
{\it Type} 1: $\sum_{\hbar} \mathcal {H}_{ii_\hbar} \geq \triangle I_i(t)$

Suppose $\phi = \sum_{\hbar} \mathcal {H''}_{ii_\hbar}(0\leq \phi\leq\triangle I_i(t))$, which represents the number of infected hosts coming from $\triangle I_i(t)$. Given a fixed $\phi$, there may be more than one permutation $H''_i = \{\mathcal {H}''_{ii_j}|j=1,\ldots,\rho\}$ for $H''_i$. The transferring likelihood estimator is
\begin{equation}
\Omega_{pu}=\sum_{\phi=0}^{\triangle I_i(t)}\sum_{\sum \mathcal {H}''_{ii_\hbar}=\phi}P_1P_2,
\label{eqP1P2}
\end{equation}
where
\begin{eqnarray}
\begin{aligned}
P_1&=
\frac{\triangle I_i(t)!}
{\prod_{\hbar} \mathcal {H}''_{ii_\hbar}!
(\triangle I_i(t)-\phi)!}\cdot\nonumber\\
&\prod_\hbar(\frac{p_\hbar}{\sum_{k=1}^{l+\rho} p_k})^{\mathcal {H}''_{ii_\hbar}}
(\frac{\sum_\ell p_\ell}
{\sum_{j=1}^{l+\rho} p_{j}})
^{\triangle I_i(t)-\phi},
\end{aligned}
\label{eqP1}
\end{eqnarray}
\begin{eqnarray}
&P_2= \nonumber\frac{(I_i(t-1) - \triangle I_i(t))!}{\prod_\hbar \mathcal {H}'_{ii_\hbar}!
(I_i(t-1) - \triangle I_i(t)-\sum_\hbar \mathcal {H}_{ii_\hbar}+\phi)!}\prod_\hbar p_\hbar^{\mathcal {H}'_{ii_\hbar}}\cdot\\
&(\sum_\ell p_\ell+\overline{p_i})^{I_i(t-1) - \triangle I_i(t)-\sum_\hbar \mathcal {H}_{ii_\hbar}+\phi}.\nonumber
\label{eqP2}
\end{eqnarray}
{\it Type} 2: $\sum_\hbar \mathcal {H}_{ii_\hbar} < \triangle I_i(t)$

Suppose $\phi = \sum_{\hbar} \mathcal {H''}_{ii_\hbar}(0\leq \phi\leq\sum_{\hbar} \mathcal {H}_{ii_\hbar})$, which represents the number of infectious hosts coming from $\triangle I_i(t)$. Given a fixed $\phi$, there may be more than one solution for $H''_i$. The transferring likelihood estimator is
\begin{equation}
\Omega_{pu}=\sum_{\phi=0}^{\sum_\hbar \mathcal {H}_{ii_\hbar}}\sum_{\sum \mathcal {H}''_{ii_\hbar}=\phi}P_1P_2,
\end{equation}
where $P_1$ and $P_2$ are the same as those in Eq.~(\ref{eqP1P2}).

According to Eq. (17), the most-likely invasion pathways for an INC $mI \mapsto nS$ can be identified as
\begin{equation}
\begin{aligned}
\hat{a}^{mInS} &= \arg \mathop{\max}\limits_{\sigma_i}~P(\sigma_i|\mathcal {E}_{mInS})\\
                &= \arg \mathop{\max}\limits_{a_i}~P(a_i|G_{mInS}).
\end{aligned}
\end{equation}

\begin{figure}[!t]
\centering
\includegraphics[width=2.5in]{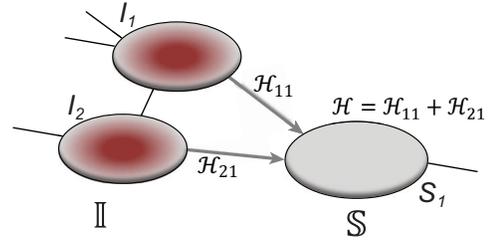}
\caption{An example of $2I\mapsto S$ invasion case. Suppose that three infected cases reach subpopulation $S_1$ simultaneously, which means
$\mathcal {H}=3$. The three possible permutations are:
$\textcircled{1} \mathcal {H}=3, \mathcal {H}_{11}=1,\mathcal {H}_{21}=2$; $\textcircled{2} \mathcal {H}=3, \mathcal {H}_{11}=2,\mathcal {H}_{21}=1$; $\textcircled{3} \mathcal {H}=3, \mathcal {H}_{11}=3,\mathcal {H}_{21}=0$.
The permutations $\textcircled{1}$ and $\textcircled{2}$ indicate the same pathways, but $\textcircled{3}$ is different.}
\label{fmerge}
\end{figure}

\noindent Note that if the first arrival infectious individuals $\mathcal {H}\geq3$, there may be multiple potential solutions corresponding to one potential pathway. For example, a $mI \mapsto S$ is illustrated in Fig.~\ref{fmerge}. In this situation, we merge the transferring likelihood of potential solutions of $mI \mapsto S$ or $mI \mapsto nS$ if they belong to the same invasion pathways, then find out the most-likely invasion pathways, which are corresponding to the maximum transferring likelihood.

According to Eq. (1) and (2), the whole invasion pathway $\mathrm{T}$ can be reconstructed chronologically by assembling all identified invasion pathway of each invasion case after identification of four classes of invasion cases. To depict the IPI algorithm explicitly, the pseudocode for our algorithm is given in Algorithm II.

\subsection{Analysis of IPI Algorithm}
Science IPI algorithm is based on hierarchical-iteration-like decomposition technique, which reduce the temporal-spatial complexity of spreading, it can handle large-scale spatial pandemic. Note that the invasion infected hosts $\mathcal {H}_i$ at EAT always are very small (generally $\leq 3$). Therefore, the computation cost of our IPI algorithm is small, and we employ the enumeration algorithm to compute each of $M$ potential permutations. In this section, we only discuss the simplest situation that one pathway only corresponds to one permitted solution in an invasion case. The situation of one pathway corresponds to multiplex potential solutions can be extended.

Denote $\pi$ the probability corresponding to the most likely pathways for a given invasion case. Thus we have
\begin{eqnarray}
\pi(\sigma) = \sup_{\sigma_i}\{P(\sigma_i|\mathcal {E})\}.
\end{eqnarray}

{\em Property 1:} Given an invasion case `$mI\mapsto S $' or `$mI\mapsto nS $', $P(\sigma_j|\mathcal {E}) = \frac{\prod_{k=1}^m \Omega}{\sum_{i=1}^M \prod_{k=1}^m\Omega}$, there must exist $P_{min}$ and $P_{max}$ satisfying
\begin{eqnarray}
P_{min} \leq \pi(\sigma) \leq P_{max}.
\end{eqnarray}
\begin{IEEEproof} Suppose that $\prod_{k=1}^mP(I_k(t),\sigma_1)\leq \ldots \leq \prod_{k=1}^mP(I_k(t),\sigma_M)$. Thus $P_{max} =
\frac{\prod_{k=1}^mP(I_k(t),\sigma_M)}{\prod_{k=1}^mP(I_k(t),\sigma_{M-1})+\prod_{k=1}^mP(I_k(t),\sigma_M)}$;
Because $\pi(\sigma) > 1/M$, let $P_{min} = max\{1/M, \frac{\prod_{k=1}^mP(I_k(t),\sigma_j)}{\prod_{k=1}^mP(I_k(t),\sigma_1)+\sum_{i=1}^M \prod_{k=1}^mP(I_k(t),\sigma_j)}\}$. We have $P_{min} \leq \pi(\sigma) \leq P_{max}$.
\end{IEEEproof}

\begin{algorithm}
\caption{Invasion Pathways Identification (IPI)}
\label{IPI}
1: Inputs: the time series of infection data $F_i(t)$ and topology of network $G(V,E)$ (including diffusion rates $p$)\\
2:  Find all invasion events via EAT data\\
3: \textbf{for} each invasion event\\
4: ~~Invasion partition to find out the $I \mapsto S$ , $I \mapsto nS$, $mI \mapsto S$ and $mI \mapsto nS$.\\
5: ~~~\textbf{for} each $mI \mapsto S$ or $mI \mapsto nS$\\
6: ~~~~~~~~~\textbf{if} it satisfy conditions of Th 1 or Th 2\\
7:  ~~~~~~~~~~~~compute the unique invasion pathway\\
8: ~~~~~~~~~\textbf{end if}\\
9:  ~~~~~~~~~\textbf{if} don't satisfy conditions of Th 1 or Th 2\\
               ~~~~~~~~~~compute the all $M$ potential solutions $\sigma_i$\\
10: ~~~~~~~~~~~compute the $P(\sigma_i|\mathcal {E}_{mIS})$ or $P(\sigma_i|\mathcal {E}_{mInS})$\\
11: ~~~~~~~~~merge the $P(\sigma_i|\mathcal {E}_{mIS})$ or $P(\sigma_i|\mathcal {E}_{mInS})$ of \\
  ~~~~~~~~~~potential solution $\sigma_i$ if they belong to same pathway\\
12: ~~~~~~~~\textbf{end if}\\
13: ~~~\textbf{end for}\\
14: ~~~~~find the maximal $a_j^{mIS}$ and $a_j^{mInS}$ invasion pathway\\
15: \textbf{end for}\\
16:  reconstruct the whole invasion pathways (T) by assembling each invasion cases chronologically\\
\end{algorithm}

\subsection{Identifiability of Invasion Pathway}
Accordingly, our IPI algorithm first decomposes the whole invasion pathways into four classes of invasion cases. Some invasion cases are easy to identify, but some are difficult. Therefore, it is important to describe how possible an invasion case can be wrongly identified. The identification extent of an invasion case relates with the absolute value of $\pi(\sigma)$ and information given by the probability vector of all potential invasion pathways. We employ the entropy to describe the information of likelihood vector, which contains the all likelihood of $M$ potential solutions/pathways of an invasion case.

{\em Definition 1 (Entropy of Transferring Likelihoods of $M$ Potential Solutions):} According to Shannon entropy, we define the normalized entropy of transferring likelihood $P(\sigma_1|\mathcal {E}),\ldots,P(\sigma_M|\mathcal {E})$ as
\begin{eqnarray}
\mathcal {S} = -\frac{1}{\log M}\sum_{i=1}^M P(\sigma_i|\mathcal {E})\log P(\sigma_i|\mathcal {E}).
\end{eqnarray}
This likelihood entropy $S$ tells the information embedded in the likelihood vector of the potential solutions of a given invasion case.

The bigger of $\pi(\sigma)$ and the smaller of entropy $\mathcal {S}$, the easier to identify the epidemic pathways for an invasion case. Define identifiability of invasion pathways to characterize the feasibility an invasion case can be identified
\begin{eqnarray}
\Pi = \pi(\sigma)(1-\mathcal {S}).
\end{eqnarray}
Although the likelihood entropies of some invasion cases are small (less than 0.5), they are still difficult to identify, because their $\pi(\sigma)$ are much less than 0.5. Therefore, identifiability $\Pi$ describes the practicability of a given $mI\mapsto S$ or $mI\mapsto nS$ better than only using $\pi(\sigma)$ or likelihoods entropy $\mathcal {S}$.
The identifiability statistically tells us why some invasion cases are easy to identify, whose $\Pi$ are more than 0.5, and why some invasion cases are difficult to identify, whose $\Pi$ are much less than 0.5.

Next we show that there exist the upper and lower boundaries of identifiability $\Pi$ for a given invasion case.

{\em Theorem 3:} Given an invasion case `$mI\mapsto S $' or `$mI\mapsto nS $', $\Pi = \pi(\sigma)(1-\mathcal {S})$ is the identifiability computed by the IPI algorithm. There exist a lower boundary $\Pi_{min}=\frac{1}{M}(1-\mathcal {S}')$ and an upper boundary $\Pi_{max}=\pi - \mathcal {S}(\pi(\sigma))$ that
\begin{eqnarray}
\Pi_{min}\leq \Pi \leq \Pi_{max},
\end{eqnarray}
where $S' = -\frac{1}{\log M}(\pi\log(\pi)+\sum\frac{1-\pi}{M-1}\log(\frac{1-\pi}{M-1}))$.

\begin{IEEEproof}
$\Pi = \pi(1-\mathcal {S})\geq\frac{1}{M}(1-\mathcal {S})\geq \frac{1}{M}(1-\mathcal {S}')$, where
$S' = -\frac{1}{\log M}(\pi\log(\pi)+\sum\frac{1-\pi}{M-1}\log(\frac{1-\pi}{M-1}))
= -\frac{1}{\log M}(\pi\log(\pi)+(1-\pi)\log(\frac{1-\pi}{M-1}))
= -\frac{1}{\log M}(\pi\log(\pi)+(1-\pi)\log(1-\pi)-(1-\pi)\log(M-1))$. According to Fano's inequality, the entropy $\mathcal {S} \leq \mathcal {S}'$.

On the other hand, we note that function $f(y)=y\log(y)$ is strictly convex. According to Jensen's inequality, $\pi \mathcal {S}(\sigma)= \pi\times (-\pi(\sigma)\log(\pi(\sigma))-\sum_{i=1}^{M-1} P(\sigma_i|\mathcal {E})\log P(\sigma_i|\mathcal {E})) \geq -(\pi(\sigma))^2\log((\pi(\sigma))^2)-\sum_{i=1}^{M-1} \pi(\sigma)P(\sigma_i|\mathcal {E})\log \pi(\sigma)P(\sigma_i|\mathcal {E})$. $\Pi = \pi(1-\mathcal {S}) = \pi- \pi \mathcal {S}\leq \pi - \mathcal {S}(\pi\sigma)$.
Therefore, $\Pi_{min}\leq \Pi \leq \Pi_{max}$, where $\Pi_{min}=\frac{1}{M}(1-\mathcal {S}')$ and $\Pi_{max}=\pi - \mathcal {S}(\pi(\sigma))$.
That completes the proof of Theorem 3.
\end{IEEEproof}

\section{Computational Experiments}\label{sec.4}

To verify the performance of our algorithm, we proceed networked metapopulation-based Monte Carlo simulation method to simulate stochastic epidemic process on the American airports network(AAN) and the Barabasi-Albert (BA) networked metapopulation.

The AAN is a highly heterogeneous network. Each node of the AAN represents an airport, the population size of which is the serving area's population of this airport. The directed traffic flow is the number of passengers through this edge/airline. The data of the AAN we are used to simulate is based on the true demography and traffic statistics~\cite{Wang:11}. We take the maximal component consisted of $404$ nodes (airports/subpopulations) of all American airports as the network size of the ANN. The average degree of the AAN is nearly $\langle k\rangle = 16$. The total population of the AAN is the $N_{total} \approx 0.243\times10^9$, which covers most of the population of the USA.

The BA network obeys heterogeneous degree distribution~\cite{Alb:02}, which holds two properties of growth and preference attachment. For a BA networked metpopulation, each node is a subpopulation containing many individuals.
The details of how to generate a BA networked metapopulation including travel rates setting is presented in Appendix B.
To test the performance of our algorithm to handle large-scale network, the subpopulations number of the BA networked metapopulation is fixed as 3000. This is nearly equal to the number of the world airports network~\cite{Barrat:04}. We fix $\langle k\rangle = 16$ as the average degree of the BA networked metapopulation. The initial population size of each subpopulation is $N_1=N_2=\cdots=N_N=6\times10^5$, and the total population is $N_{total} = 6\times10^5\times3000= 1.8\times10^9$, which covers most of the active travelers of the world.

\begin{figure}[!t]
\centering
\includegraphics[width=3.3in]{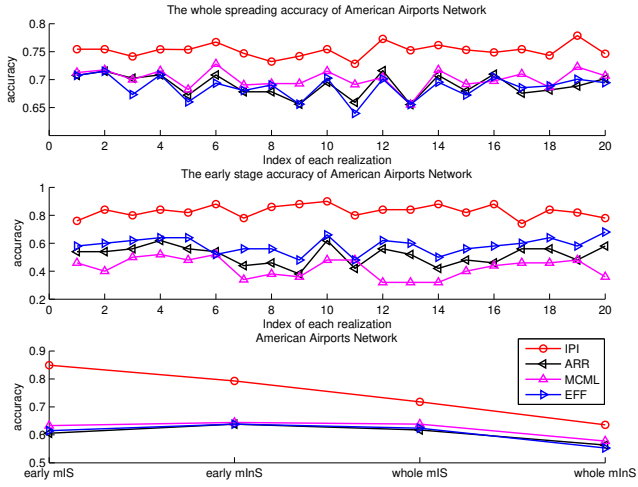}
\caption{The top and middle figures show the identified accuracy for the whole and early stage (before appearance of the first 50 infected subpopulations) invasion pathways for twenty independent spreading realizations on the AAN. The bottom shows accumulative identified accuracy of invasion cases ($mI\mapsto S $ and $mI\mapsto nS $) for the early stage and the whole invasion pathways on the AAN.}
\label{fig_sim}
\end{figure}

\subsection{Networked Metapopulation-based Monte Carlo Simulation Method to Simulate Stochastic Epidemic Process}
 At the beginning, we assume only one subpopulation is seeded as infected and others are susceptible. Thus $I_{1}(0)=5, I_i(0)=0(i=2,\cdots,N)$. We record and update each individual's state(i.e., susceptible or infected) at each time step. At each $\Delta t$ ($\Delta t$ is defined as the unit time from $t-1$ to $t$), the transmission rate $\beta$ and diffusion rate $p_{ij}$ are converted into probabilities. The rules of individuals reaction and diffusion process in $\Delta t$ are as follows:

\noindent (1) Reaction Process: Individuals which are in the same subpopulation are homogeneously mixing. Each susceptible individual (in subpopulation $i$) becomes infected with probability $\beta\frac{I_i(t)}{N_i}$. Therefore, the average number of newly added infected individuals is $\beta\frac{(N_i(t)-I_i(t))I_i}{N_i}$, but the simulation results fluctuate from one realization to another. The reaction process is simulated by binomial distribution.

\noindent (2) Diffusion Process: After reaction, the diffusion process of individuals between different subpopulation posterior to the reaction process is taken into account. Each individual from subpopulation $i$ migrates to the neighboring subpopulation $j$ with probability $p_{ij}$.
The average {number of} new infectious travelers from subpopulation $i$ to $j$ is $p_{ij}I_i(t)$. The diffusion process is simulated by binomial distribution or multinomial distribution.

\subsection{Numerical Results}

We compared our IPI algorithm with three heuristic algorithms that generate the shortest path tree or minimum spanning tree of the metapopulation
networks. i) The ARR (average-arrival-time-based shortest path tree)~\cite{Gau:08}: The minimum distance path from subpopulation $i$ to subpopulation $j$ over all possible paths is generated in terms of mean first arrival time. Thus the average-arrival-time-based shortest path tree is constructed
by assembling all shortest paths from the seed subpopulation to other subpopulations of the whole network. ii) The EFF (effective-distance-based most probable paths tree)\cite{Dirk:13} methods: From subpopulation $i$ to subpopulation $j$, the effective distance $D_{ij}$ is defined as the minimum of
the sum of effective lengths along the arbitrary legs of the path. The set of shortest paths to all subpopulations from seed subpopulation $i$ constitutes a shortest path tree. ii) The MCML (the Monte-Carlo-Maximum-Likelihood-based most likely epidemic invasion tree)\cite{Duy:09}: To produce a most likely infection tree, they constructed the minimum spanning tree from the seed subpopulation to minimize the distance. Some recent works~\cite{Wan:14}-\cite{Yang:14} uses machine learning or genetic algorithms to infer transmission networks from surveillance data. Because of the
distinction in model assumptions and conditions, we do not perform comparison with them.

We consider to access the identification accuracy for the inferred invasion pathways. This accuracy is defined by the ratio between the number of corrected identified invasion pathways by each method and the number of true invasion pathways, respectively. We also compute the accuracy of accumulative invasion cases of $mI\mapsto S $ and $mI\mapsto nS $. This accuracy is defined by the ratio between the number of corrected identified invasion pathways by each method in this class of invasion case and the number of true invasion pathways in this classes of invasion case. Additionally, we investigate the identification accuracy of early stage of a global pandemic spreading, which is important to help understand how to predict and control the prevalence of epidemics.

\begin{figure}[!t]
\centering
\includegraphics[width=3.35in]{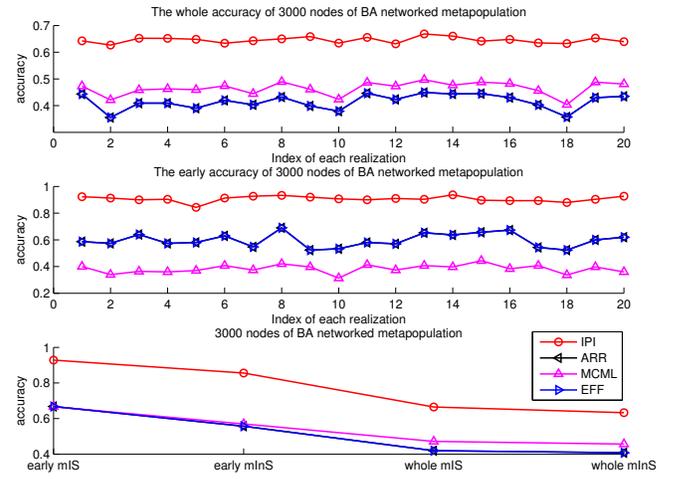}
\caption{The top and middle figures show the identified accuracy for the whole and early stage invasion pathways for twenty independent spreading realizations on 3000 subpopulations of the BA networked metapopulation. The bottom shows accumulative identified accuracy of $mI\mapsto S $ and $mI\mapsto nS $ for the early stage (the first 300 infected subpopulations) and the whole invasion pathways on 3000 subpopulations of the BA networked metapopulation.}
\label{fig_sim}
\end{figure}

In the top and middle of Figure 5, we observe the whole identification accuracy and the early-stage identification accuracy.  The bottom of Figure 5 shows the early and whole accumulative identification accuracy of $mI\mapsto S $ and $mI\mapsto nS $ through twenty independent realizations on the AAN  for each algorithm, respectively. The simulation results show our algorithm is outperformance, which indicates heterogeneity of structure of the AAN plays an important role.

Figure 6 shows the results of the BA networked metapopulation with 3000 subpopulations, the top of which presents the identification accuracy of whole invasion pathway for each realization of the four algorithms. while the middle of Figure 6 shows the identification accuracy of early stage invasion pathway for each realization. The bottom shows accumulative identified accuracy of $mI\mapsto S $ and $mI\mapsto nS $ of twenty realizations for four alorithms respectively. The simulation results indicate that our algorithm can handle a large scale networked metapopulation with robust performance.
Note that the performance of the ARR for the BA networked metapopulation is the same as that of the EFF, because our parameter $C$ is a constant in the diffusion model (see Appendix B).

The numerical results suggest that networks with different topologies yield different identification performances, which indicate an identification algorithm should embed in both the effects of spreading and topology. Our algorithm takes into account both the heterogeneity of epidemics (the number of infected individuals) and the network topology (diffusion flows).

\begin{figure}[!t]
\centering
\includegraphics[width=3.5in]{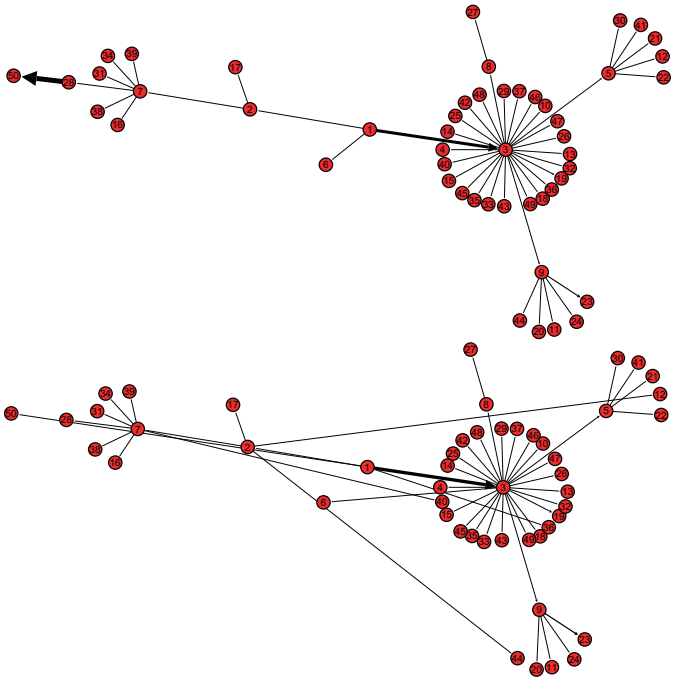}
\caption{Illustration of the actual invasion pathways and the most likely identified invasion pathways, in a given realization, during the early stage
(before the appearance of 50 infected subpopulations) on the AAN. Subpopulation 1 is the seed.}
\label{fig_sim}
\end{figure}

We finally test the identifiability of an invasion case. Figure 8 shows the entropy and identifiability of wrongly identified $mI \mapsto S$ of twenty realizations on the AAN. The smaller the identifiability of an invasion case is, the easier it is prone to be wrongly identified. The identifiability depicts the wrongly identified $mI \mapsto S$ more reasonable than the likelihoods entropy.
It indicates that identifiability $\Pi$ has a better performance to distinguish whether an invasion case is difficult to identify or not than using the likelihoods entropy.

\section{Conclusion and discussion}\label{sec.5}

To conclude, we have proposed an identification framework as the so called IPI algorithm to explore the problem of inferring invasion pathway for a pandemic outbreak. We first anatomize the whole invasion pathway into four classes of invasion cases at each epidemic arrival time. Then we identify four classes of invasion cases, and reconstruct the whole invasion pathway from the source subpopulation of a spreading process. We introduce the concept of {\it identifiability} to quantitatively analyze the difficulty level that an invasion case can be identified. The simulation results on the American Airports Network (AAN) and large-scale BA networked metapopulation have demonstrated our algorithm held a robust performance to identify the spatial invasion pathway, especially for the early stage of an epidemic.
We conjecture the proposed IPI algorithm framework can extend to the problems of virus diffusion in computer network, human to human's epidemic contact network, and the reaction dynamics may extend to the SIR or SIS dynamics.\\

\begin{figure}[!t]
\centering
\includegraphics[width=2.35in]{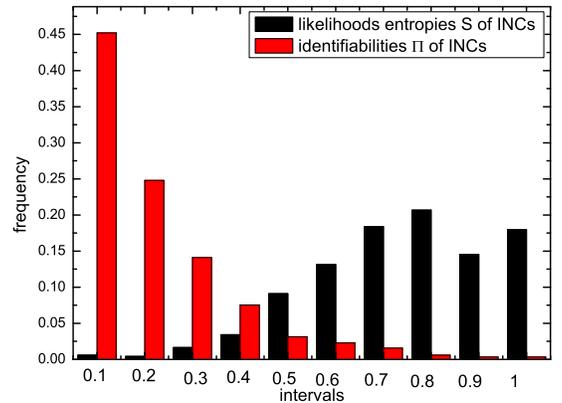}
\caption{Statistics analysis of the likelihoods entropy and identifiability of wrongly identified $mI \mapsto S$ of 20 realizations of epidemic spreading  on the AAN.}
\label{fig_sim}
\end{figure}

\section*{appendix A}

\begin{center} \small
MOBILITY OPERATOR
\end{center}

We discuss the individual {\em mobility operator}.
Due to the presence of stochasticity and independence of individual mobility, the number of successful transform of individuals
between or among adjacent subpopulations is quantified by a binomial or a multinomial process, respectively. If the
focal subpopulation $i$ only has one neighboring subpopulation $j$, the number of individuals in a given compartment
$\mathcal {X}$ ($\mathcal {X}\in \{S,I\}$ and $\sum_{\mathcal {X}} \mathcal {X}_i=N_i$) transferred from $i$ to $j$
per unit time, $\mathcal {T}_{ij}(\mathcal {U}_i)$, is generated from a binomial distribution with probability $p$
representing the diffusion rate and the number of trials $\mathcal {U}_i$, i.e.,
\begin{equation}
Binomial(\mathcal {T}_{ij},\mathcal {U}_i,p)=\frac{\mathcal {U}_i!}{\mathcal {T}_{ij}!(\mathcal {U}_i-\mathcal {T}_{ij})!}p^{\mathcal {T}_{ij}}{(1-p)}^{(\mathcal {U}_i-\mathcal {T}_{ij})}.
\end{equation}
If the focal subpopulation $i$ has multiple neighboring subpopulations $j_1,j_2,...,j_k$, with $k$ representing $i$'s
degree, the numbers of individuals in a given compartment $\mathcal {U}$ transferred from $i$ to $j_1,j_2,...,j_k$ are
generated from a multinomial distribution with probabilities $p_{ij_1},p_{ij_2},...,p_{ij_k}$ ($p_{ij_1}+p_{ij_2}+...+p_{ij_k}=p$) representing the diffusion rates on the edges emanated from
subpopulation $i$ and the number of trails $\mathcal {U}_i$, i.e.,
\begin{equation}
\begin{aligned}
&Multinominal(\{\mathcal {T}_{ij_{\ell}}\},\mathcal {U}_i,\{p_{ij_{\ell}}\})=&\\
&\frac{\mathcal {U}_i!}{\prod_{\ell}\mathcal {T}_{ij_{\ell}}!(\mathcal {U}_i-\sum_{\ell}\mathcal {T}_{ij_{\ell}})!}
(\prod_{\ell}p_{ij_{\ell}}^{\mathcal {T}_{ij_{\ell}}})(1-\sum_{\ell}p_{ij_{\ell}})^{(\mathcal {U}_i-\sum_{\ell}\mathcal {T}_{ij_{\ell}})},&\label{eq.2}
\end{aligned}
\end{equation}
where $\ell\in[1,k]$.

\section*{appendix B}\small
\begin{center}
A GENERIC DIFFUSION MODEL TO GENERATE \\A BARABASI-ALBERT METAPOPULATION NETWORK
\end{center}

We develop a general diffusion model to generate a BA metapopulation network in Section V, which characterizes the human mobility pattern on the empirical statistical rules of air transportation networks.

The diffusion rate from subpopulation $i$ to $j$ is $p_{ij} = \frac{w_{ij}}{N_i}$, where $w_{ij}$ denotes the traffic flow from subpopulation $i$ to $j$. These empirical statistical rules are verified in the air transportation network~\cite{Barrat:04}: $\langle p_{ij} \rangle \sim (k_ik_j)^{\theta'},{\theta'}=0.5\pm0.1;T\sim k^{\beta'},\beta'\simeq1.5\pm0.1; N\sim T^{\lambda'}(T=\sum_l w_{jl}), \lambda' \simeq 0.5$.

All the above empirical formulas relate to node's degree $k$. To generate an artificial transportation network, we introduce a generic diffusion model to determine the diffusion rate
\begin{equation}
p_{ij} = \frac{b_{ij}k_j^{\theta}}{\sum_l b_{il}k_{l}^{\theta}}C,
\end{equation}
where $b_{ij}$ stands for the elements of the adjacency matrix ($b_{ij}=1$ if $i$ connects to $j$, and $b_{ij}=0$ otherwise),
 $C$ is a constant, and $\theta$ is a variable parameter. We assume that parameter $\theta$ follows the Gaussian distribution
$\theta \sim N(\hat{\theta},\delta^2) = \frac{1}{\sqrt{2\pi}\delta}exp(-\frac{(\hat{\theta}-\theta)^2}{2\delta^2})$.
Based on the empirical rule of $T\sim k^{\beta'},\beta'\simeq1.5\pm0.1$, where $\beta$ is approximately linear with $\theta$, the least squares estimation is employed to evaluate parameters $\hat{\theta}$ and $\delta^2$ if we set the initial population of each node and constant $C$. Then, for a given BA network, we get a BA networked metapopulation in which real statistic information is embedded by using the above method.

\end{document}